\documentclass{elsart}

\usepackage{graphicx}
\usepackage{subfigure}
\usepackage{amsmath}

\input epsf

\begin{document}

\setcounter{footnote}{0}
\setcounter{figure}{0}

\begin{frontmatter}
\title{
Studies of correlations between $D$ and ${\overline D}$ mesons
in high energy photoproduction
}

$\textrm{The~FOCUS~Collaboration}^\star$
\thanks{See \textrm{http://www-focus.fnal.gov/authors.html} for
additional author information.}

\author[ucd]{J.~M.~Link},
\author[ucd]{P.~M.~Yager},
\author[cbpf]{J.~C.~Anjos},
\author[cbpf]{I.~Bediaga},
\author[cbpf]{C.~G\"obel},
\author[cbpf]{J.~Magnin},
\author[cbpf]{A.~Massafferri},
\author[cbpf]{J.~M.~de~Miranda},
\author[cbpf]{I.~M.~Pepe},
\author[cbpf]{E.~Polycarpo},
% ...

\address[ucd]{University of California, Davis, CA 95616}
\address[cbpf]{Centro Brasileiro de Pesquisas F\'isicas, Rio de Janeiro, RJ, Brasil}
\address[cinv]{CINVESTAV, 07000 M\'exico City, DF, Mexico}
\address[cu]{University of Colorado, Boulder, CO 80309}
\address[fnal]{Fermi National Accelerator Laboratory, Batavia, IL 60510}
\address[fras]{Laboratori Nazionali di Frascati dell'INFN, Frascati, Italy I-00044}
\address[ugj]{University of Guanajuato, 37150 Leon, Guanajuato, Mexico}
\address[ui]{University of Illinois, Urbana-Champaign, IL 61801}
\address[iu]{Indiana University, Bloomington, IN 47405}
\address[korea]{Korea University, Seoul, Korea 136-701}
\address[kp]{Kyungpook National University, Taegu, Korea 702-701}
\address[milan]{INFN and University of Milano, Milano, Italy}
\address[nc]{University of North Carolina, Asheville, NC 28804}
\address[pavia]{Dipartimento di Fisica Nucleare e Teorica and INFN, Pavia, Italy}
\address[pr]{University of Puerto Rico, Mayaguez, PR 00681}
\address[sc]{University of South Carolina, Columbia, SC 29208}
\address[ut]{University of Tennessee, Knoxville, TN 37996}
\address[vu]{Vanderbilt University, Nashville, TN 37235}
\address[wisc]{University of Wisconsin, Madison, WI 53706}

\begin{abstract}

Studies of $D{\overline D}$ correlations for a large sample of events
containing fully and partially reconstructed pairs of charmed $D$ mesons
recorded by the Fermilab photoproduction experiment FOCUS (FNAL-E831) are
presented. Correlations between $D$ and ${\overline D}$ mesons are used to
study heavy quark production dynamics. We present
results for fully and partially reconstructed charm pairs
and comparisons to a recent version of \textsc{Pythia} with default
parameter settings. We also comment on the production of $\psi(3770)$ in our
data.

\end{abstract}

\end{frontmatter}

\section{Introduction}

Heavy quark production continues to present itself as a challenge to our
understanding of the strong interaction. While Quantum Chromodynamics (QCD)
provides a theoretical framework for our understanding and perturbative QCD can
be applied to some aspects of heavy quark production, other aspects remain
elusive and cannot be described without including a variety of non-perturbative
effects. This is especially true for charm production, where perturbative QCD
calculations involve large uncertainties and non-perturbative effects play a
significant role in modeling physical observables. Until we achieve a
fundamental understanding of the strong interaction, accurate models that are
able to reproduce properties of the strong interaction---such as heavy quark
production---are crucial for our understanding of this fundamental force.

In this paper, we present new results from FOCUS (FNAL-E831) on charm-pair
correlations between $D$ and ${\overline D}$ mesons.
Charm-pair correlations
have received considerable theoretical
%attention~\cite{frix98}\cite{frix94}\cite{frix94x}\cite{man93}\cite{man92}\cite{nas},
attention~\cite{frix98,frix94,frix94x,man93,man92,nas},
and have been studied in both
%hadroproduction~\cite{e791}\cite{wa92}\cite{e653}\cite{na32}\cite{wa75}\cite{na27}
hadroproduction~\cite{e791,wa92,e653,na32,wa75,na27}
and
photoproduction~\cite{na14,e687}
experiments. We present our photoproduction
results by comparing data distributions to predictions from a recent version of
a Monte Carlo based on the Lund Model~\cite{pyth6203},
which includes
non-perturbative effects that have been shown to be important in charm
production. We select default settings for charm photoproduction in the Monte
Carlo to facilitate comparisons with theoretical predictions and results from
other experiments.

% references from Frixione paper for experimental (and theoretical?) work
% references from E791 paper and thesis

\section{Experimental method}

The data for our studies of $D{\overline D}$ correlations were recorded by the
FOCUS experiment during the 1996--1997 fixed-target run at the Fermi National
Accelerator Laboratory. The experiment ran with a photon beam\footnote{The
photon beam was produced from the bremsstrahlung of secondary electrons and
positrons with an endpoint energy of $\approx$~300 GeV. The average photon
energy for the recorded data was $\approx$~180 GeV with a width of
$\approx$~50 GeV.}
and a spectrometer that was upgraded from a previous
photoproduction experiment, E687~\cite{e687_spec}. The FOCUS spectrometer had a
target that consisted of four BeO target elements for most of the recorded
data\footnote{Early in the run a few different targets were used, and
less than 5$\%$ of the charm-pair
data were recorded with Be (instead of BeO) target elements.}.
A vertex detector, which was
located in the target region, had a total of 16 planes of silicon strip
detectors. Four of the planes were interleaved with the BeO target elements,
and 12 were located downstream of the target. Tracks that were reconstructed in
the vertex detector were linked to particle tracks that were found in five
multiwire proportional chambers. Particle momenta were determined by measuring
the deflection of tracks in two analysis magnets of opposite polarity, and
particle identification was accomplished using measurements from three
multicell threshold \v{C}erenkov counters, details of which
are described elsewhere~\cite{cerenkov}.

\setcounter{figure}{0}

\begin{figure}[!htp]
\centering
\subfigure[]
{\includegraphics[width=0.4\textwidth,clip=true]{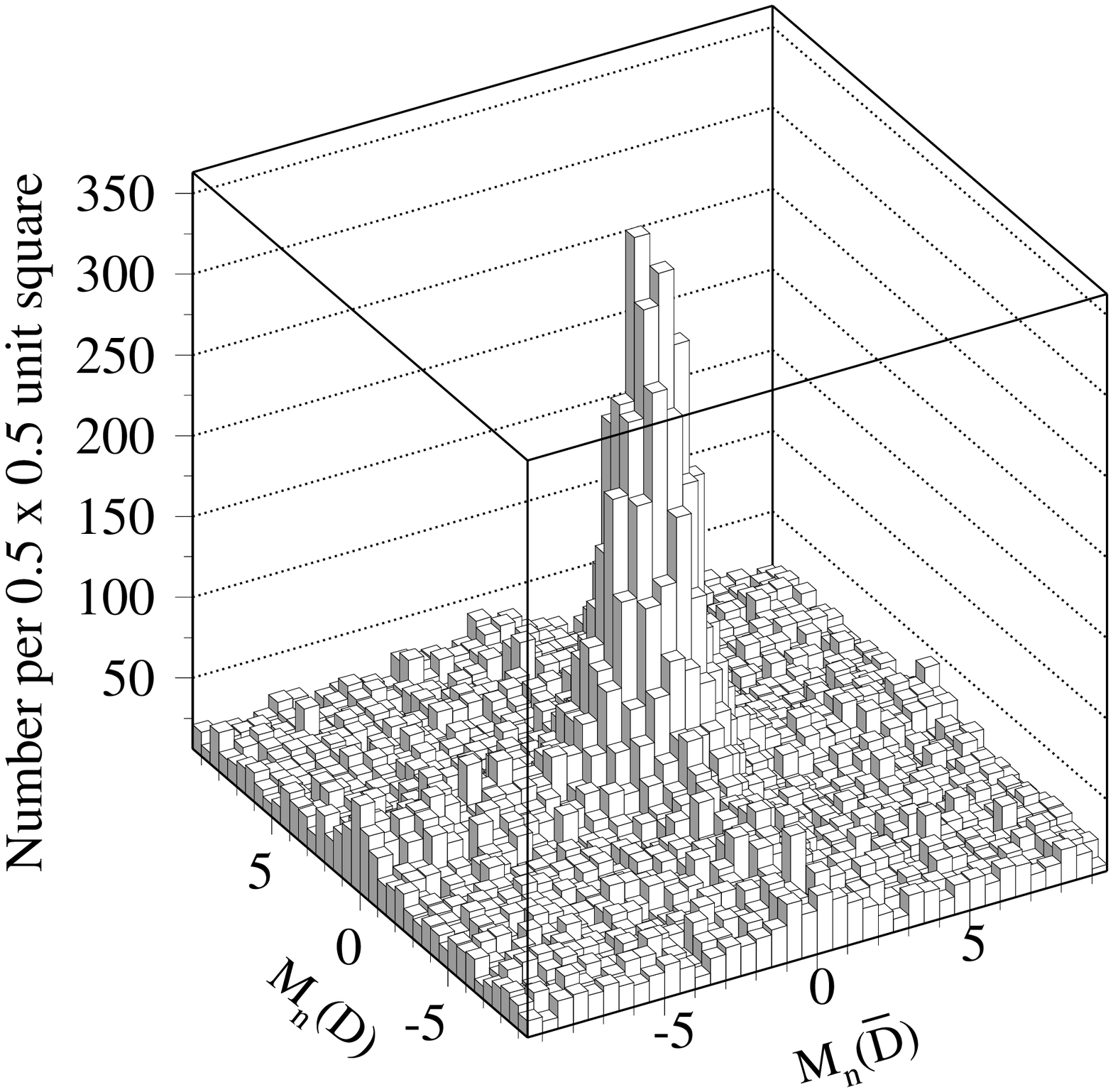}}
\subfigure[]
{\includegraphics[width=0.4\textwidth,clip=true]{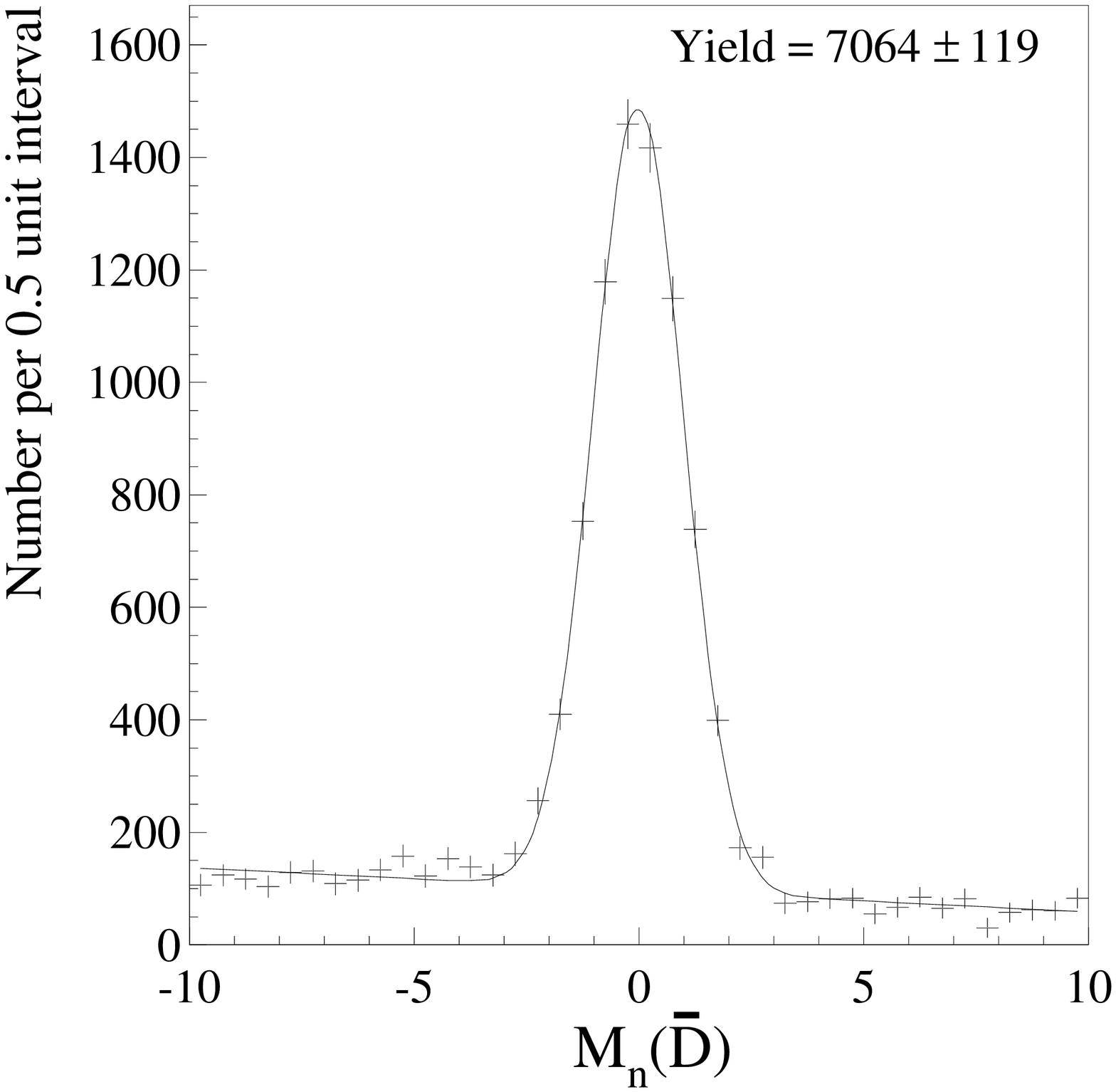}} \\
\subfigure[]
{\includegraphics[width=0.4\textwidth,clip=true]{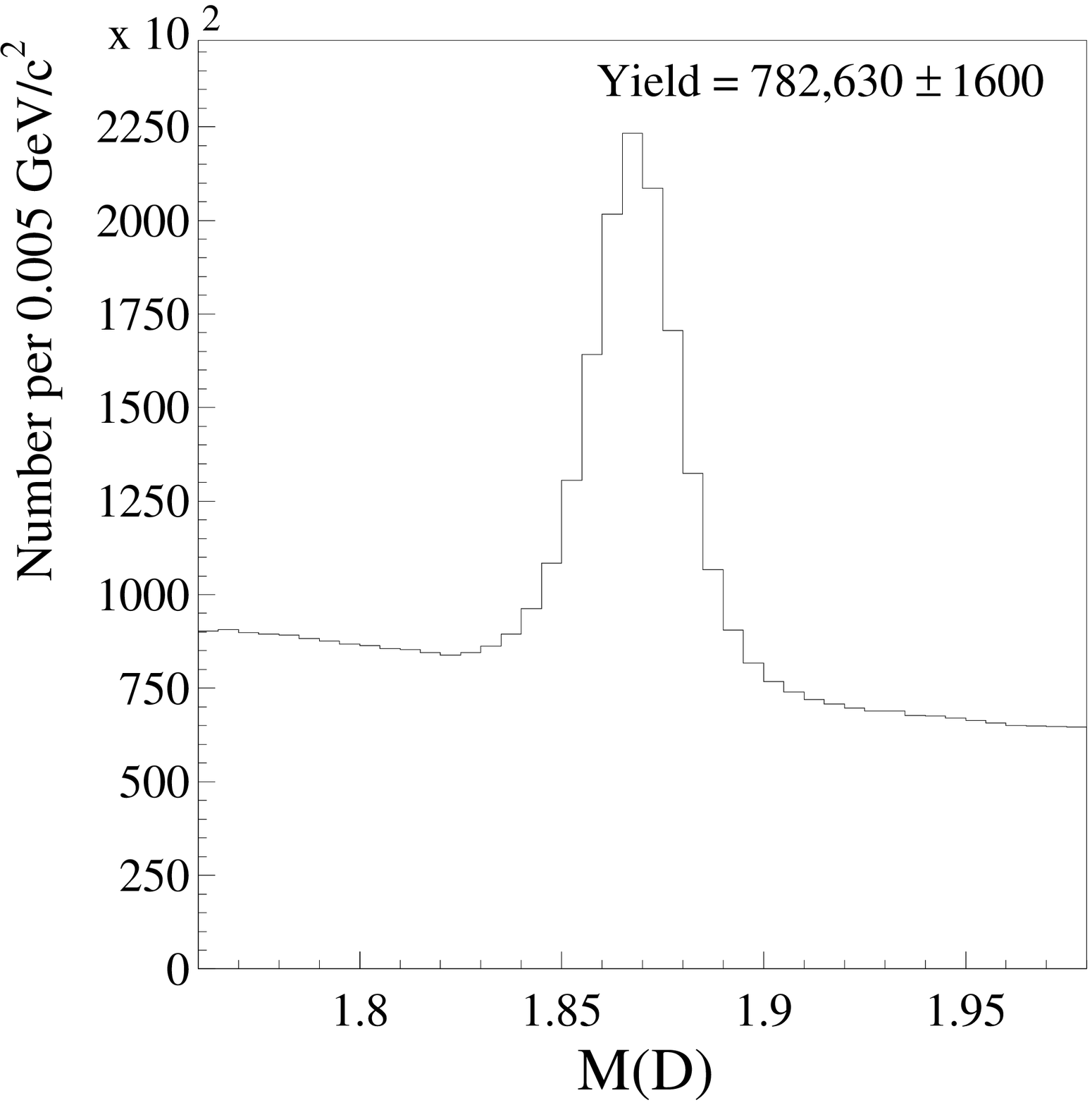}}
\subfigure[]
{\includegraphics[width=0.4\textwidth,clip=true]{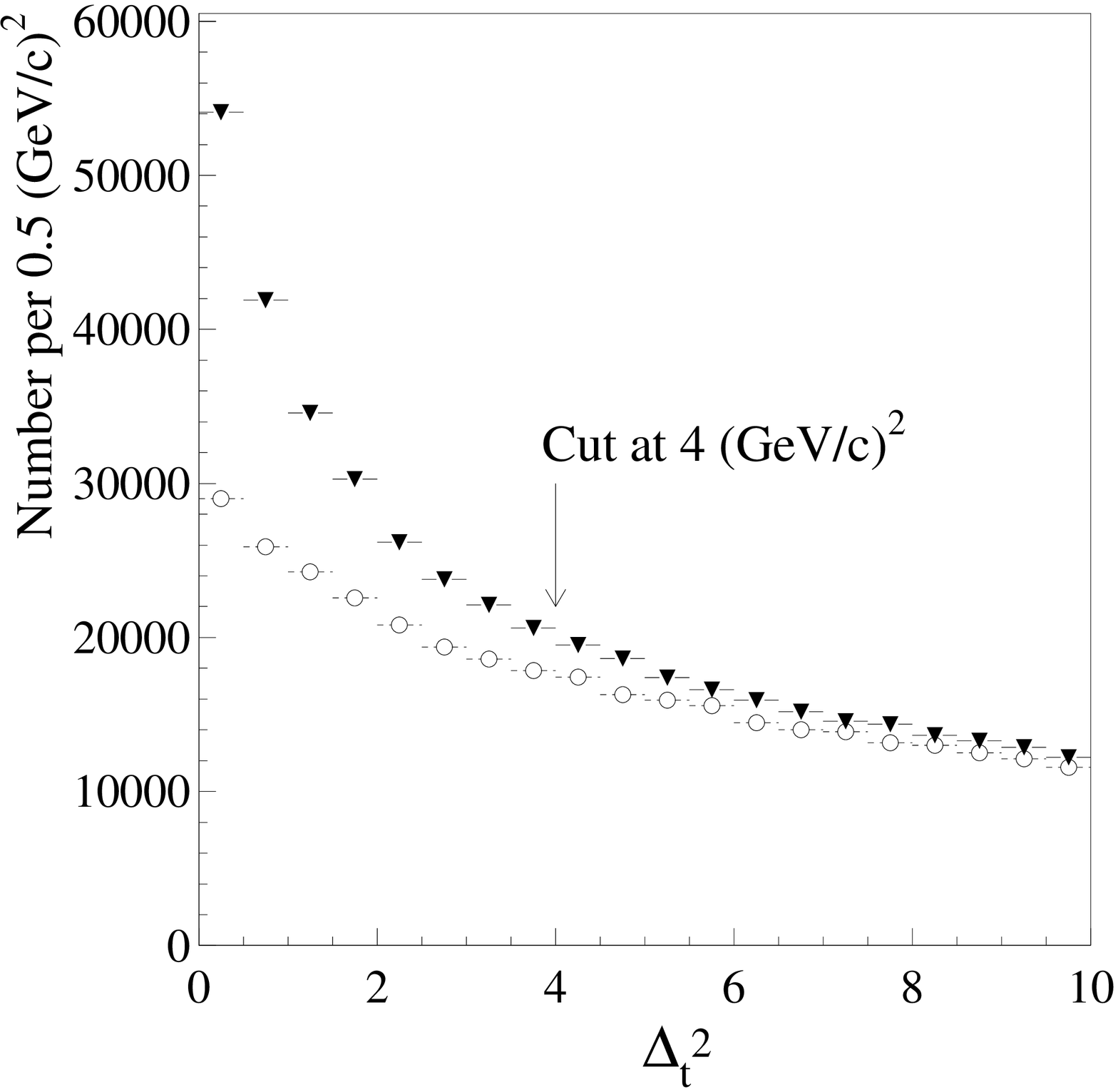}}
\label{lego-and-fit}
\caption{
(a) Normalized $D$ invariant mass vs. normalized ${\overline D}$ invariant mass
distribution, and (b) a fit to the normalized ${\overline D}$ invariant mass
after sideband subtraction (described in the text).
(c) Invariant mass of the recoil $D$ in the partially reconstructed charm-pair
sample (the mass of charged $D$ candidates is lowered by
3.74 $\textrm{MeV}/c^2$ to match the $D^0$ mass distribution). The yield is a sum of
individual yields for the three decay modes.
(d) $\Delta_t^2$ distributions for right-sign (filled triangles) and wrong-sign
(open circles) combinations for partially reconstructed charm pairs.
}
\end{figure}

Here we describe, for the first time, the candidate-driven algorithm that was
used to collect a large sample of $\approx$~7000 pairs of fully reconstructed
charmed mesons. 
The sample consists of pairs of $D$ mesons:
$D^+D^-$, $D^+{\overline D}^0$, $D^0D^-$, and $D^0{\overline D}^0$.
For
this paper, we considered the decay modes $D^0\!\rightarrow\!~K^-\pi^+$,
$D^+\!\rightarrow\!~K^-\pi^+\pi^+$,  $D^0\!\rightarrow\!~K^-\pi^+\pi^+\pi^-$, and
charged-conjugate modes.  The algorithm considered all combinations of two,
three, and four charged tracks to find a combination that could be associated
with the decay of a single $D$ meson, and a second combination of tracks that
could be associated with a second $D$ decay vertex in the same event. The
successful reconstruction of two $D$ vertex candidates was followed by
the reconstruction of a primary interaction vertex, particle identification
cuts, and detachment cuts for the $D$ vertices relative to the primary vertex.
The goal was to achieve low background levels for each decay mode using a
minimum number of cuts.

The first step of the candidate-driven algorithm considers all \emph{pairs} of
two-, three-, and four-track combinations in an event. Each combination of
tracks represents a possible $D$ decay. For each track the algorithm considers
all possible combinations of charged \emph{K} or \emph{$\pi$} assignments such
that the assignments are consistent with the decay of a charged
or neutral $D$ meson. A particular combination of tracks and the associated
particle assignments is referred to as a $D$ candidate. The mass of each $D$
candidate is calculated using the measured track momenta, and is required to
fall within a wide range of 1.6--2.4 $\textrm{GeV}/c^2$. To select events with
a $D$ and a ${\overline D}$, the kaons for the two $D$ candidates
are required to have opposite charge.

\begin{figure}[!htp]
\centering
\subfigure[]
{\includegraphics[width=0.4\textwidth,clip=true]{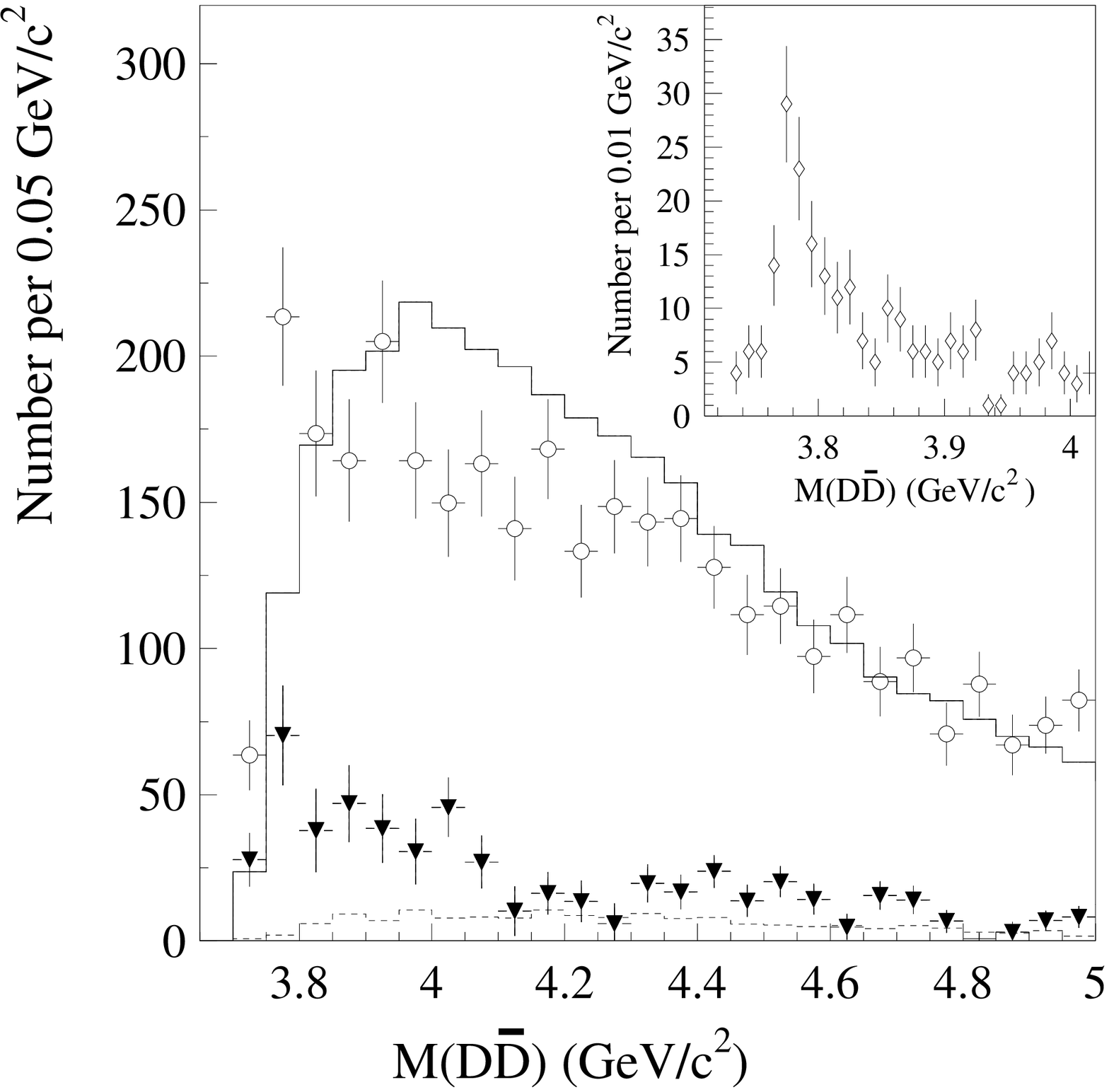}}
\subfigure[]
{\includegraphics[width=0.4\textwidth,clip=true]{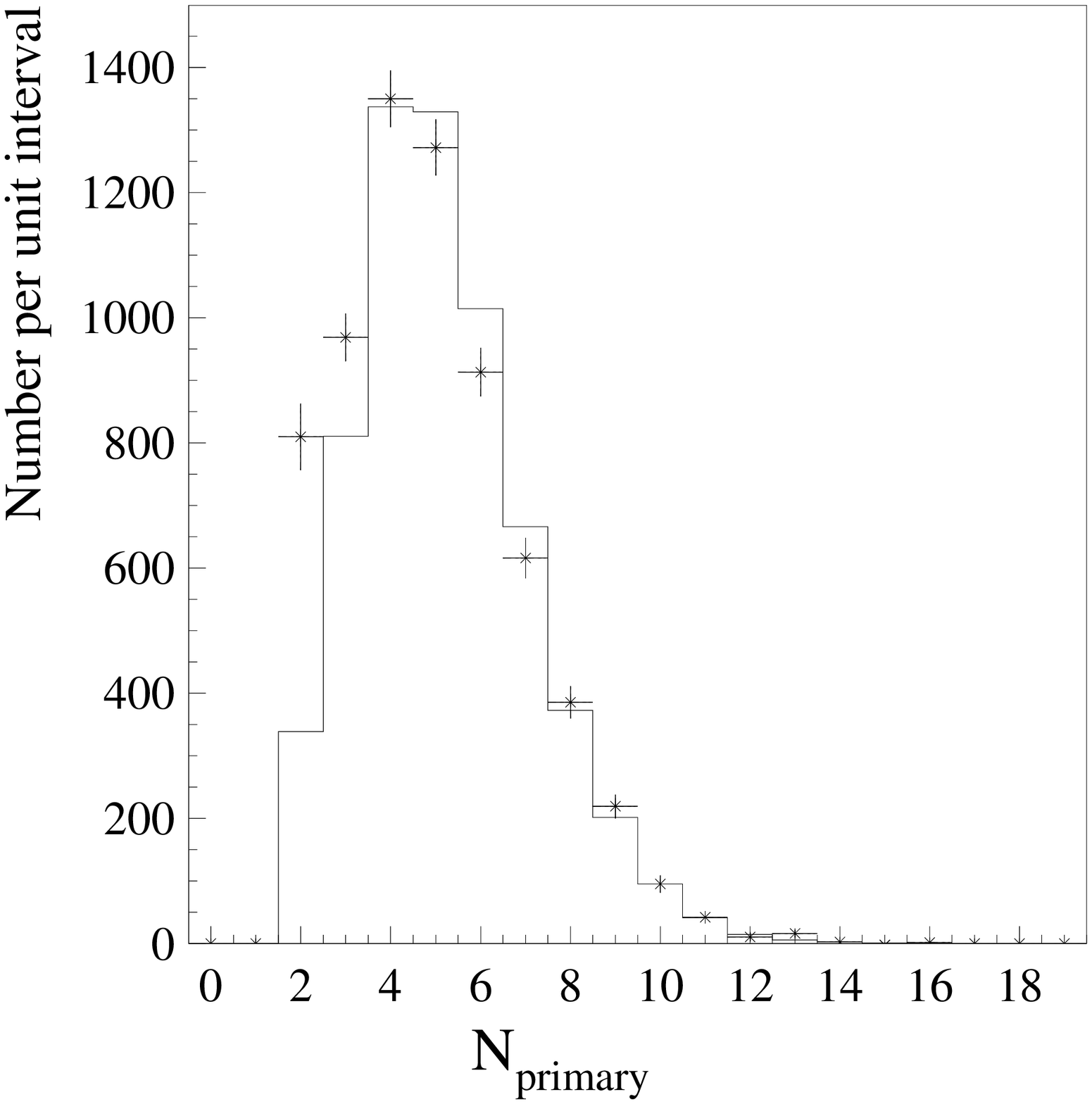}} \\
\subfigure[]
{\includegraphics[width=0.4\textwidth,clip=true]{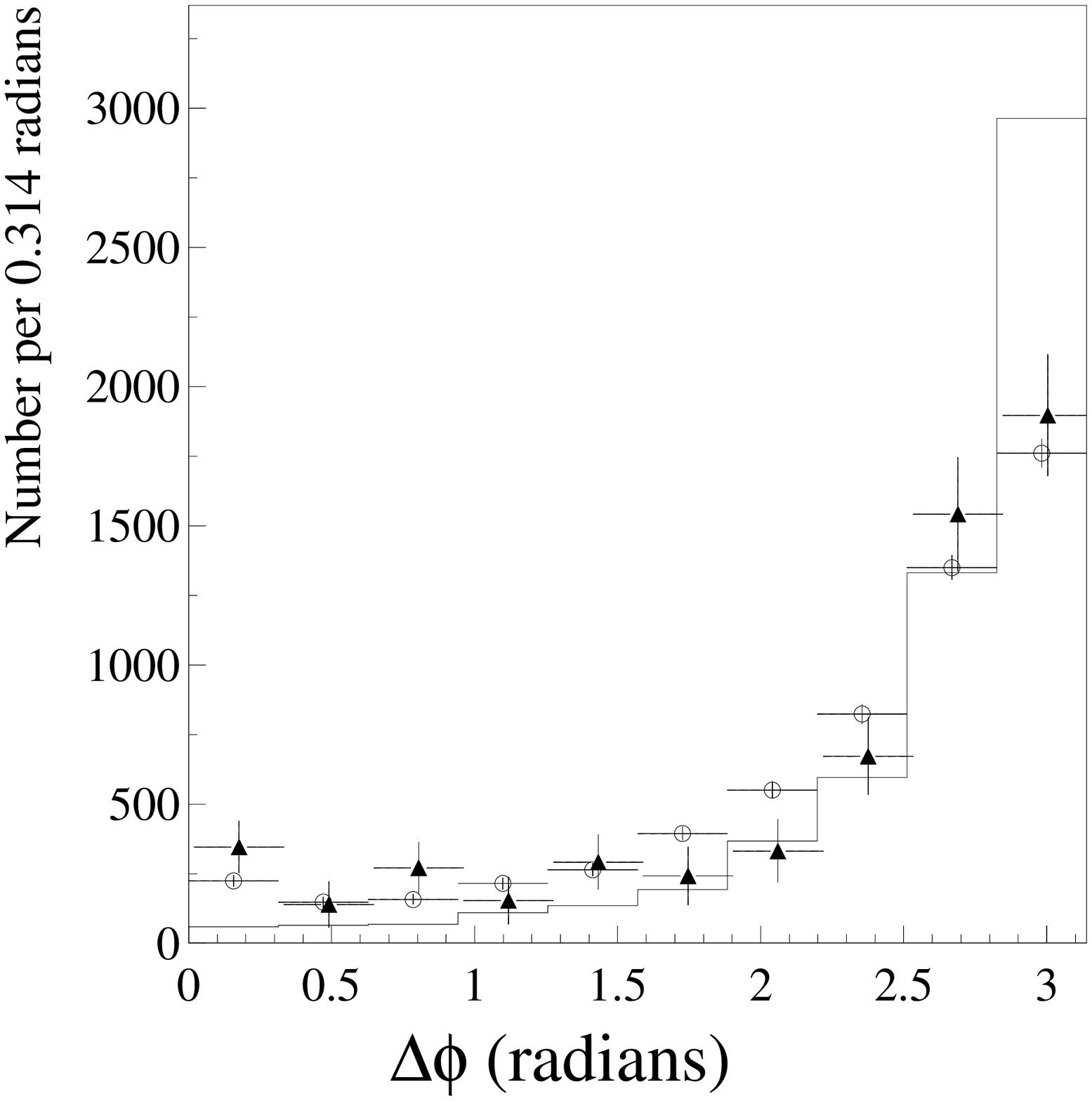}}
\subfigure[]
{\includegraphics[width=0.4\textwidth,clip=true]{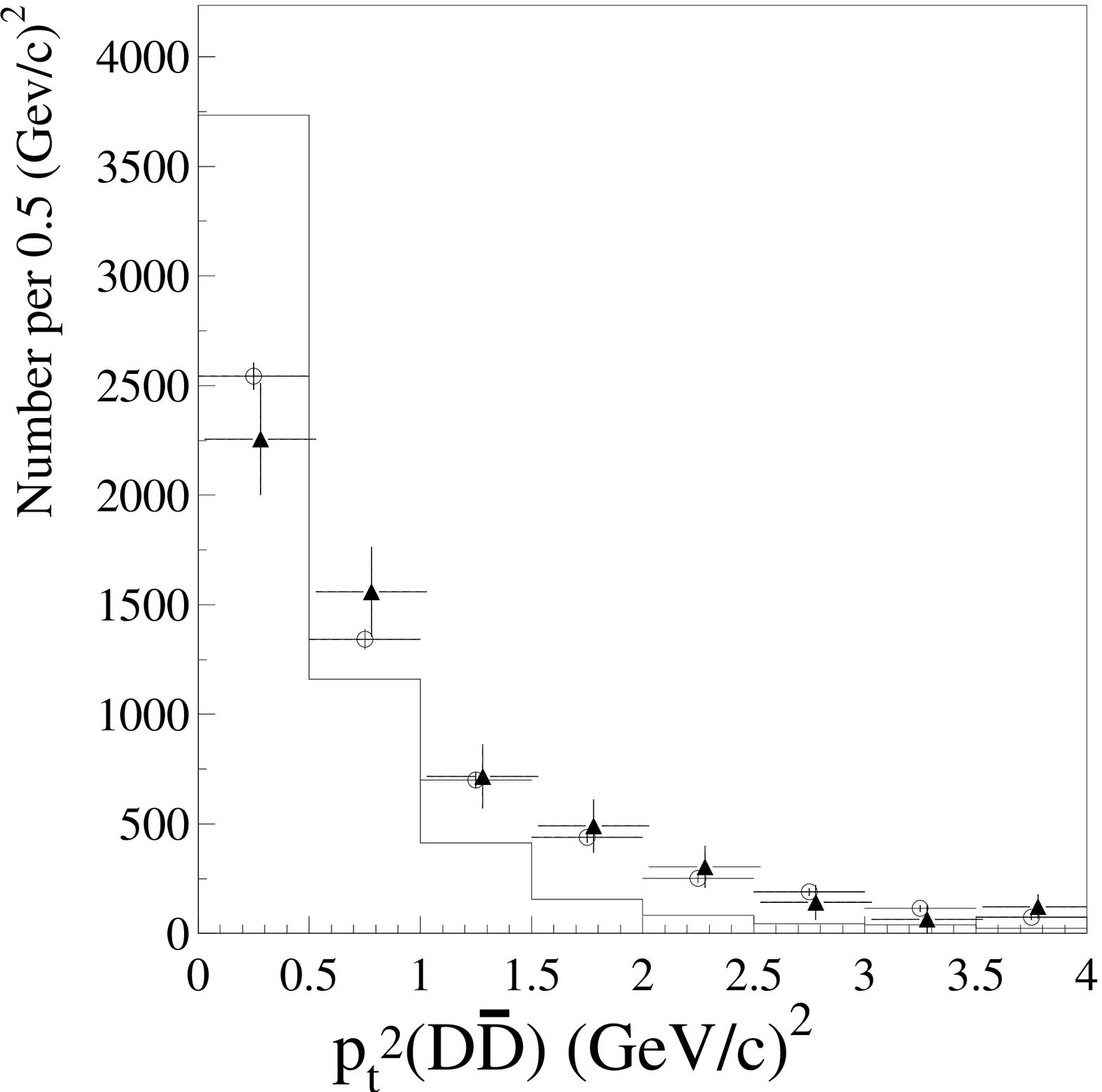}}
\label{figure2}
\caption{
(a) Invariant $D{\overline D}$ mass for $D^+D^-$ and $D^0{\overline D}^0$ mass
combinations for background-subtracted FOCUS data (open circles), \textsc{Pythia}~6.203
(solid line), FOCUS
data with $N_{\textrm{primary}}=2$ cut (filled triangles), and \textsc{Pythia}~6.203 with
$N_{\textrm{primary}}=2$ cut (dashed line). The inset shows the invariant
$D{\overline D}$ mass
that we obtain after applying additional cuts, such as
cuts that remove events with energy deposited in
the electromagnetic calorimeters. (b) Number of tracks assigned
to the primary vertex for background-subtracted FOCUS data (data points with
error bars) and
\textsc{Pythia}~6.203 (solid line) normalized to the number of $D{\overline D}$ pairs
in data with
$N_{\textrm{primary}}>2$. (c) $\Delta\phi$ and (d) $p_t^2$ of the
$D{\overline D}$ pair
for background-subtracted FOCUS data with $N_{\textrm{primary}}>2$ (open
circles), E687 data (filled triangles with
offset to show error bars) normalized to FOCUS data, and \textsc{Pythia}~5.6
(solid line).
}
\end{figure}

The second step is vertex reconstruction. The goal is to find a pair of
$D$-decay vertices that can be associated with a primary interaction vertex,
and to find all other tracks in the event that can be associated with that
primary vertex. This part of the algorithm starts by performing a vertex fit
for each $D$ candidate. The tracks for each $D$ candidate are required to form
a vertex with confidence level greater than 1$\%$. Pairs of $D$ candidates that
satisfy the confidence level cut are subjected to two additional vertex cuts.
The first cut requires that the momentum vectors of the two candidates
intersect with a confidence level greater than 1$\%$. The second cut rejects
background by rejecting pairs of $D$ candidates for which the reconstructed
daughter tracks for both $D$ candidates form a \emph{single} vertex with
confidence level greater than 0.1$\%$. This rejects background events in which
tracks for both candidates all come from a common vertex. The final phase of
the vertex reconstruction treats the two $D$ candidates as \emph{seed tracks}
to find the primary vertex. Vertex fits are performed by including the two seed
tracks as well as combinations of all other tracks in the event. As many tracks
as possible are added to the primary vertex as long as the confidence level is
greater than 1$\%$.

Pairs of $D$ candidates that survive the vertex reconstruction are subjected to
particle-identification cuts, which are based on measurements from three
multicell threshold \v{C}erenkov counters. The \v{C}erenkov
algorithm~\cite{cerenkov} calculates four likelihoods that correspond to the
four hypotheses (electron, pion, kaon, proton) that are considered for each
charged track. The algorithm produces a $\chi^2$-like variable $W_i = -2 \ln
(\mathrm{likelihood})$, where $i$ is the index used to represent each
hypothesis. For the kaon in each $D$ candidate, we require that the kaon
hypothesis is favored over the pion hypothesis by more than a factor of
exp(0.5) by requiring $W_{\pi} - W_K > 1.0$. For the pions in each $D$
candidate we apply a pion consistency cut, which requires that no particle
hypothesis is favored over the pion hypothesis with a $\Delta W = W_\pi -
W_{\textrm{min}}$ greater than $5$, where $W_{\textrm{min}}$ is the $W_i$ with
the smallest value.

After applying particle-identification cuts, we impose cuts based on the
significance of detachment ($\ell/\sigma_\ell$) between each $D$ candidate and
the primary vertex. We calculate $\ell/\sigma_\ell$ by using the measured value
of $\ell$, the distance between the $D$ decay vertex and the primary vertex,
and dividing by the associated error $\sigma_\ell$. The cuts for
$\ell/\sigma_\ell$ range from $\ell/\sigma_\ell>1$ to $\ell/\sigma_\ell>4$
depending on the decay mode, whether the $D$-decay vertex is located between
target elements (for which background levels are low) or in target material,
and whether a $D$ candidate can be associated with a $D^*$ decay.

\begin{figure}[!htp]
\centering
\subfigure[]
{\includegraphics[width=0.4\textwidth,clip=true]{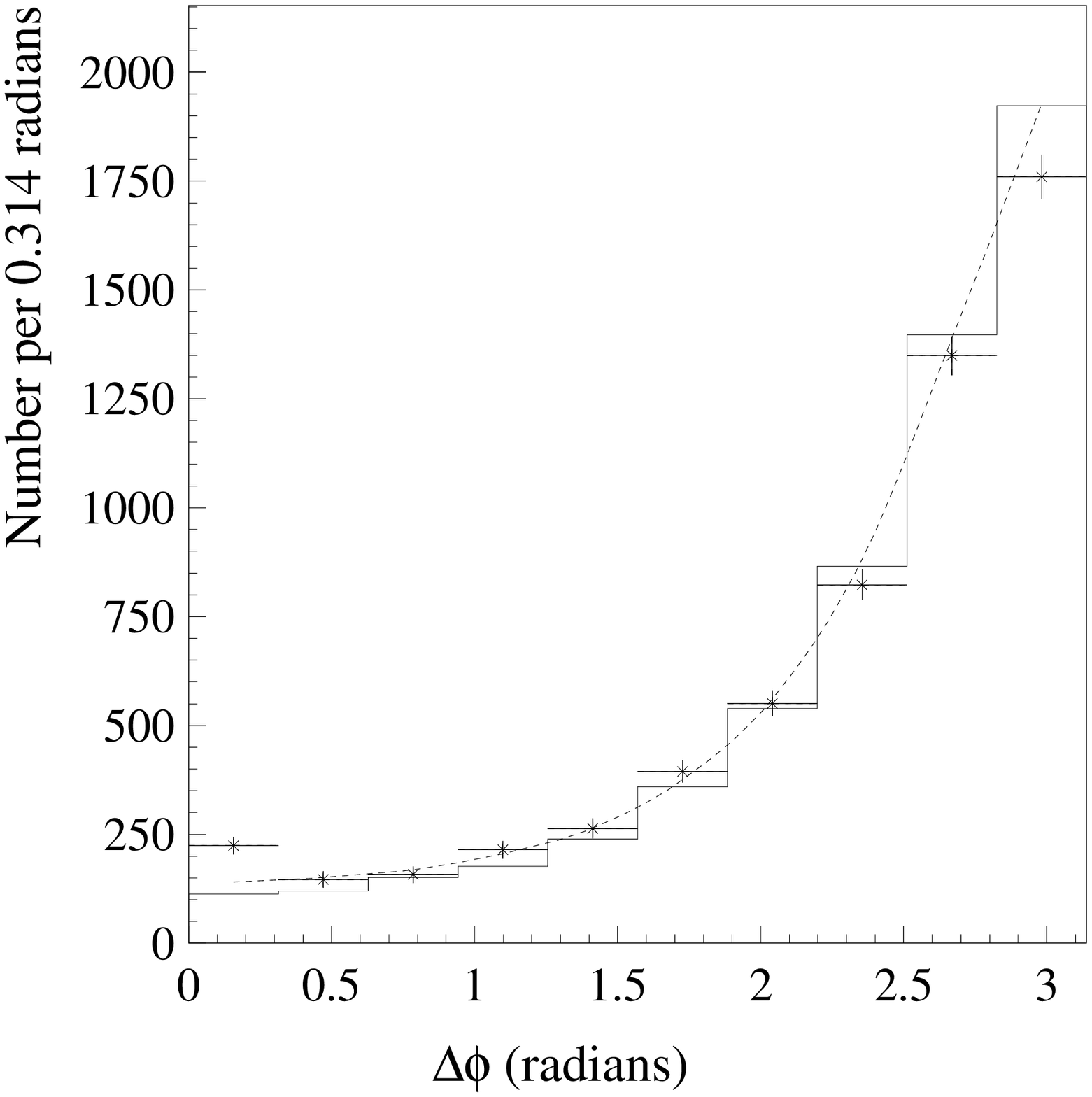}}
\subfigure[]
{\includegraphics[width=0.4\textwidth,clip=true]{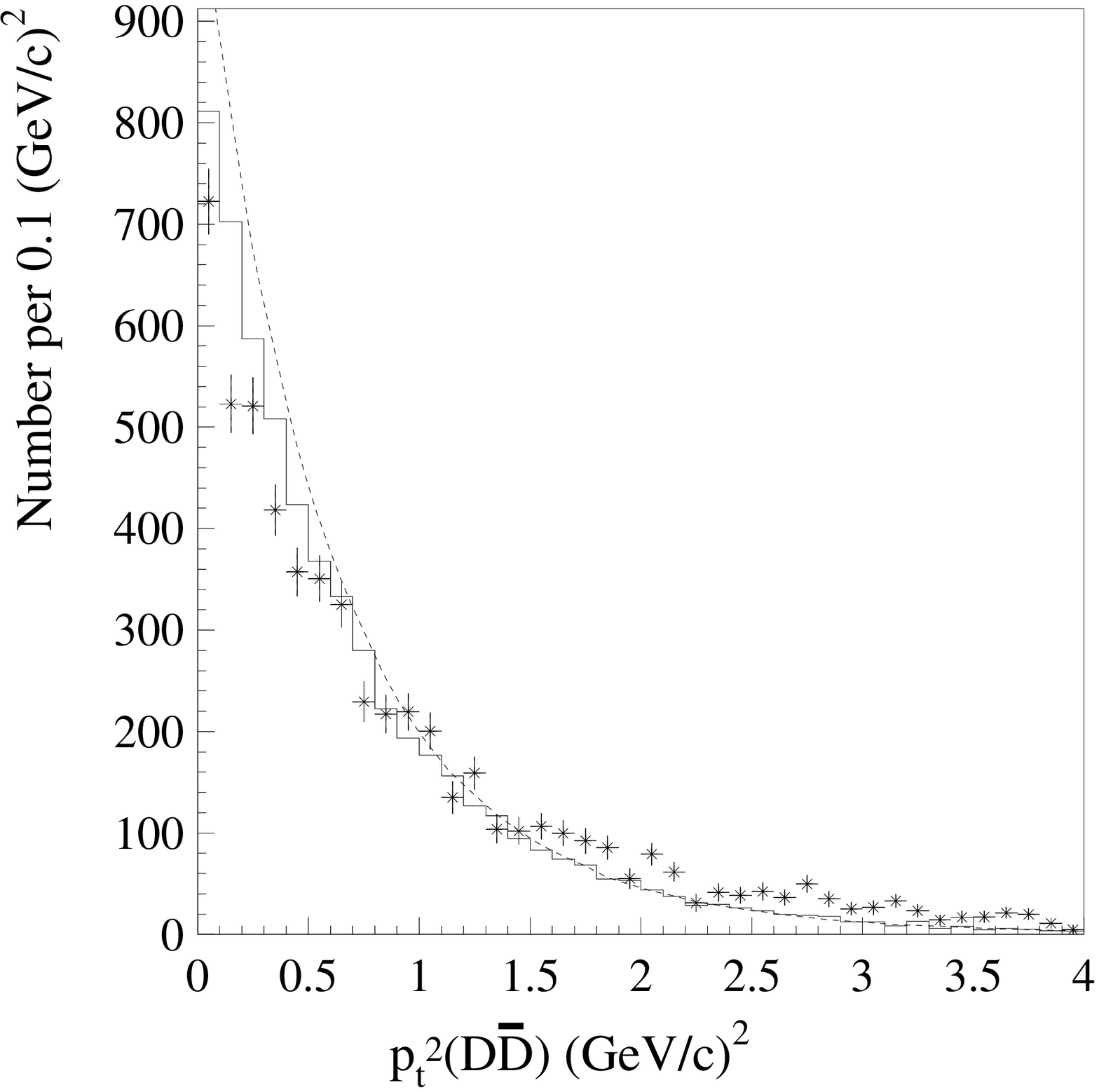}} \\
\subfigure[]
{\includegraphics[width=0.4\textwidth,clip=true]{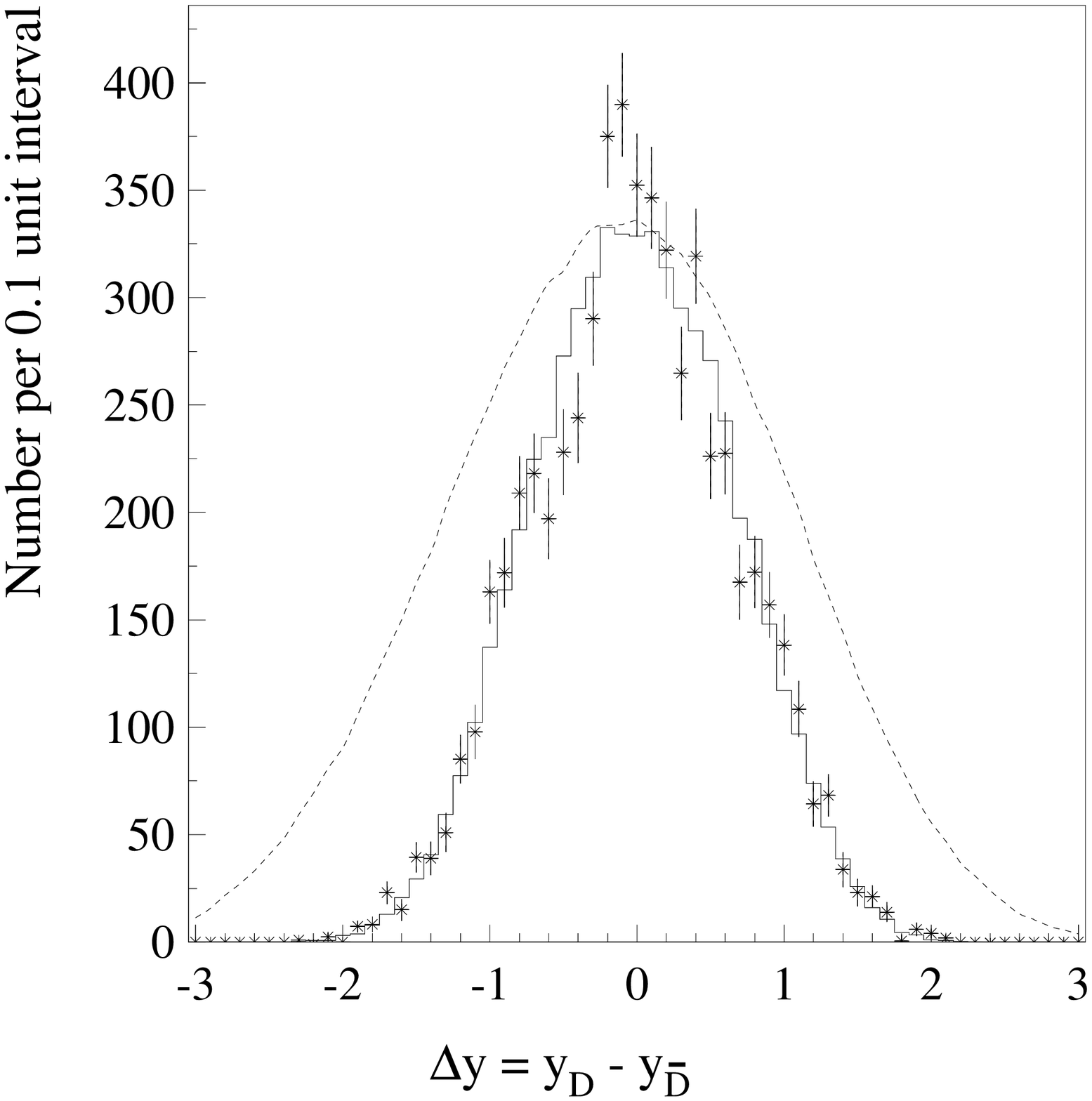}}
\subfigure[]
{\includegraphics[width=0.4\textwidth,clip=true]{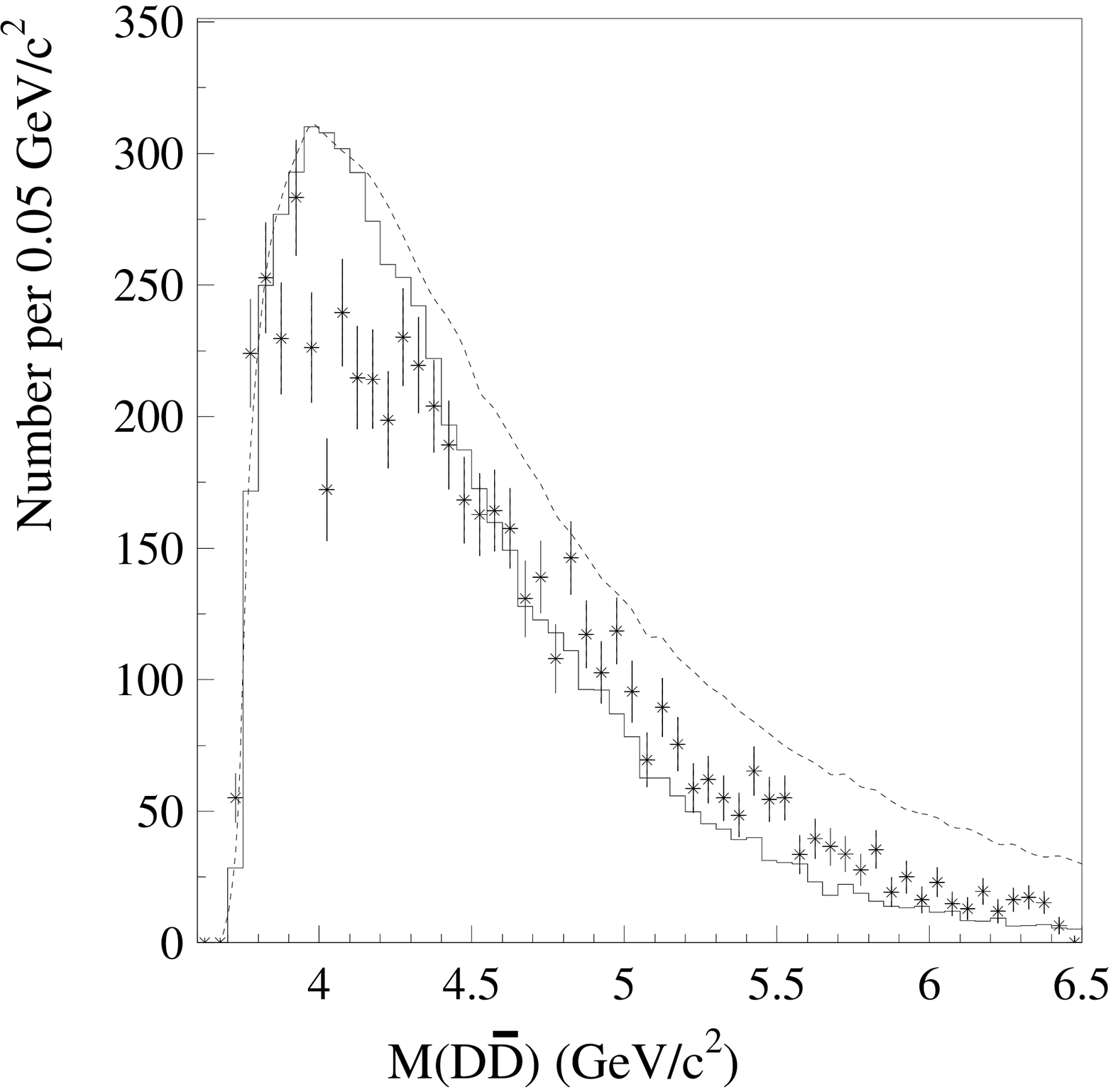}}
\label{figure3}
\caption{
Correlations for fully reconstructed $D{\overline D}$ pairs with
$N_{\textrm{primary}}>2$:
(a) $\Delta\phi$, (b) $p_t^2$ of the $D{\overline D}$ pair, 
(c) rapidity difference ($y_D - y_{\overline D}$), and
(d) invariant $D{\overline D}$ mass for background-subtracted FOCUS data
(data points with error bars), \textsc{Pythia} 6.203 after detector simulation and
data analysis cuts (solid line), and \textsc{Pythia} 6.203 parent distributions without
acceptance or resolution effects (dashed line with arbitrary normalization).
}
\end{figure}

Figure~1a shows the $D{\overline D}$ signal that we obtain after all of the
aforementioned cuts have been applied to the FOCUS data.
Figure~1a shows the normalized $D$ invariant mass\footnote{The normalized
mass, $M_n(D) = \Delta M/\sigma_M$, is defined as the difference between the
reconstructed mass and the central value of the $D^+$ or $D^0$ mass distribution
divided by the reconstructed-mass error $\sigma_M$, which is calculated for each
$D$ candidate.}
$M_n(D)$ opposite the
normalized ${\overline D}$ invariant mass $M_n({\overline D})$. Figure~1b shows a
Gaussian fit to $M_n({\overline D})$ over a linear background after applying a
background subtraction procedure that is used to determine the number of pairs
of charmed $D$ mesons in the FOCUS data. The procedure consists of performing a
sideband subtraction and fit for one normalized mass distribution by selecting
entries in the signal and sideband regions of the other normalized mass
distribution. In Fig.~1b we plot $M_n({\overline D})$ by assigning unit weight
to $D$ candidates with a reconstructed mass in the signal region ($\pm~2\sigma$
about the central value of the $D^+$ or $D^0$ mass of the candidate), and a
weight of $-^1/_2$ to candidates with mass in the two 4--8$\sigma$
sideband regions\footnote{An equivalent approach to determine the $D{\overline
D}$ yield is a fit to $M_n(D)$ after selecting signal and sideband regions for
$M_n({\overline D})$. Using this approach we obtain a $D{\overline D}$ yield of
$7126~\pm~120$ , which is consistent with the yield mentioned in the text.}.
The $D{\overline D}$ yield that we obtain from our fit is $7064~\pm~119$
(statistical error).

In addition to our study of correlations between pairs of fully reconstructed
$D$ mesons, we study correlations between two $D$ mesons where one $D$ is fully
reconstructed and the other is kinematically tagged by a slow pion coming from
the decay $D^{*+}\!\rightarrow\!~\pi^+D^0$. In these decays, the $D^0$ need not be
reconstructed, and therefore we refer to this sample of charmed $D$ mesons as
partially reconstructed charm pairs\footnote{The partially reconstructed
sample consists of
$D^{*+}D^-$, $D^{*+}{\overline D}^0$, $D^0D^{*-}$, and $D^+D^{*-}$ pairs.
}.
The reason for including this sample in
our studies of correlations is that charm-pair correlations can be
studied over a larger kinematic range compared to the fully reconstructed
sample.

For partially reconstructed charm pair events~\cite{e687} we begin by
considering all two-, three-, and four-track combinations for the fully
reconstructed $D$ (recoil $D$) in an event. We consider the decay modes
$D^0\!\rightarrow\!~K^-\pi^+$, $D^+\!\rightarrow\!~K^-\pi^+\pi^+$, 
$D^0\!\rightarrow\!~K^-\pi^+\pi^+\pi^-$, and charged-conjugate modes. A
candidate-driven algorithm uses the recoil $D$ candidates to find the primary
vertex, requiring the vertex confidence level to be greater than 1$\%$. The
same \v{C}erenkov particle identification criteria used for the fully
reconstructed charm-pair sample (see above) are applied to the recoil $D$
candidates. However, a more restrictive detachment cut of $\ell/\sigma_\ell>5$
is applied to all three decay modes in the partially reconstructed charm-pair
sample.
Figure~1c shows the invariant mass distribution, which
includes
all three decay modes,
with a total of $782\,630~\pm~1600$ candidates satisfying the selection criteria.

The next step in the analysis treats each track that is assigned to the primary
vertex (excluding the recoil $D$) as a slow-pion candidate from the decay
$D^{*+}\!\rightarrow\!~\pi^+D^0$. The momentum of the track is multiplied by 13.8
to approximate the momentum of the $D^{*+}$\footnote{Due to the low $Q$ value
of the $D^*$ decay, the momentum of the soft pion approximates the momentum of
the $D^*$ when multiplied by the inverse of its energy fraction, which is
$\approx$~13.8.
}.
If the charge of the
slow pion is the same as the charge of the kaon from the recoil $D$, then the
combination of the slow pion and recoil $D$ is designated as a
\emph{right-sign} combination. Otherwise, it is a \emph{wrong-sign}
combination. This assignment of right- and wrong-sign combinations is used
for background subtraction.

A double subtraction method is used to reduce backgrounds. First, to handle
non-charm background, a sideband subtraction is applied to recoil $D$
candidates. A Gaussian fit is applied to the invariant mass distribution for
each of the three decay channels. Entries in the 4--8$\sigma$ sideband regions
are subtracted from those in the $\pm~2\sigma$ peak region by using a weight
factor of $-^1/_2$.
Second, the assignment of right- and wrong-sign combinations is used to
subtract wrong-sign background from right-sign combinations. To avoid
distortion of the wrong-sign background we \emph{exclude} all slow-pion
candidates
that can be associated with a $D^*$ decay involving the recoil $D$. This
\emph{anti-$D^*$} cut is imposed by excluding combinations of slow pions
and recoil $D$-mesons that have a mass difference, $m(D^*)-m(D)$,
in the range 0.142--0.149 $\textrm{GeV}/c^2$.
To further
enhance the selection procedure, a maximum cut of 4 (GeV/$c$)$^2$ is applied to
         ${\Delta}_t^2 = (p_x^{(r)}+ 13.8* p_x^{(\pi)})^2
                       + (p_y^{(r)}+ 13.8* p_y^{(\pi)})^2$, 
where $p_x^{(r)}, p_y^{(r)}$ and $p_x^{(\pi)}, p_y^{(\pi)}$ are transverse
momentum components of the recoil $D$ and slow pion, respectively. This cut
enhances the selection of signal
since genuine events balance ${\Delta}_t^2$ (see reference~\cite{e687}
for more details). This is shown in Fig.~1d, which shows a prominent excess of
right-sign combinations close to ${\Delta}_t^2=0$ compared to the wrong-sign
background. After applying the double subtraction and the ${\Delta}_t^2$ cut,
we obtain a sample of $75\,160~\pm~1040$ partially-reconstructed charm
pairs.

\section{$D{\overline D}$ correlations}

For our study of correlations between pairs of fully reconstructed $D$ mesons,
we compare FOCUS data to predictions from a Monte Carlo based on the Lund
Model. The Monte Carlo consists of a \textsc{Pythia} 6.203~\cite{pyth6203} generator
with default settings, and detector simulation algorithms for the FOCUS
apparatus. The Monte Carlo generator produces charm events using a tree-level
photon-gluon fusion process applied to beam photons and target nucleons. We use
default options for charm photoproduction in the generator (instead of using a
Monte Carlo tuned to match our data) to facilitate comparisons with theoretical
predictions and results from other experiments. In this paper, we also compare
our results to previously published charm photoproduction results from
experiment E687~\cite{e687}.

\begin{figure}[!htp]
\centering
\subfigure[]
{\includegraphics[width=0.4\textwidth,clip=true]{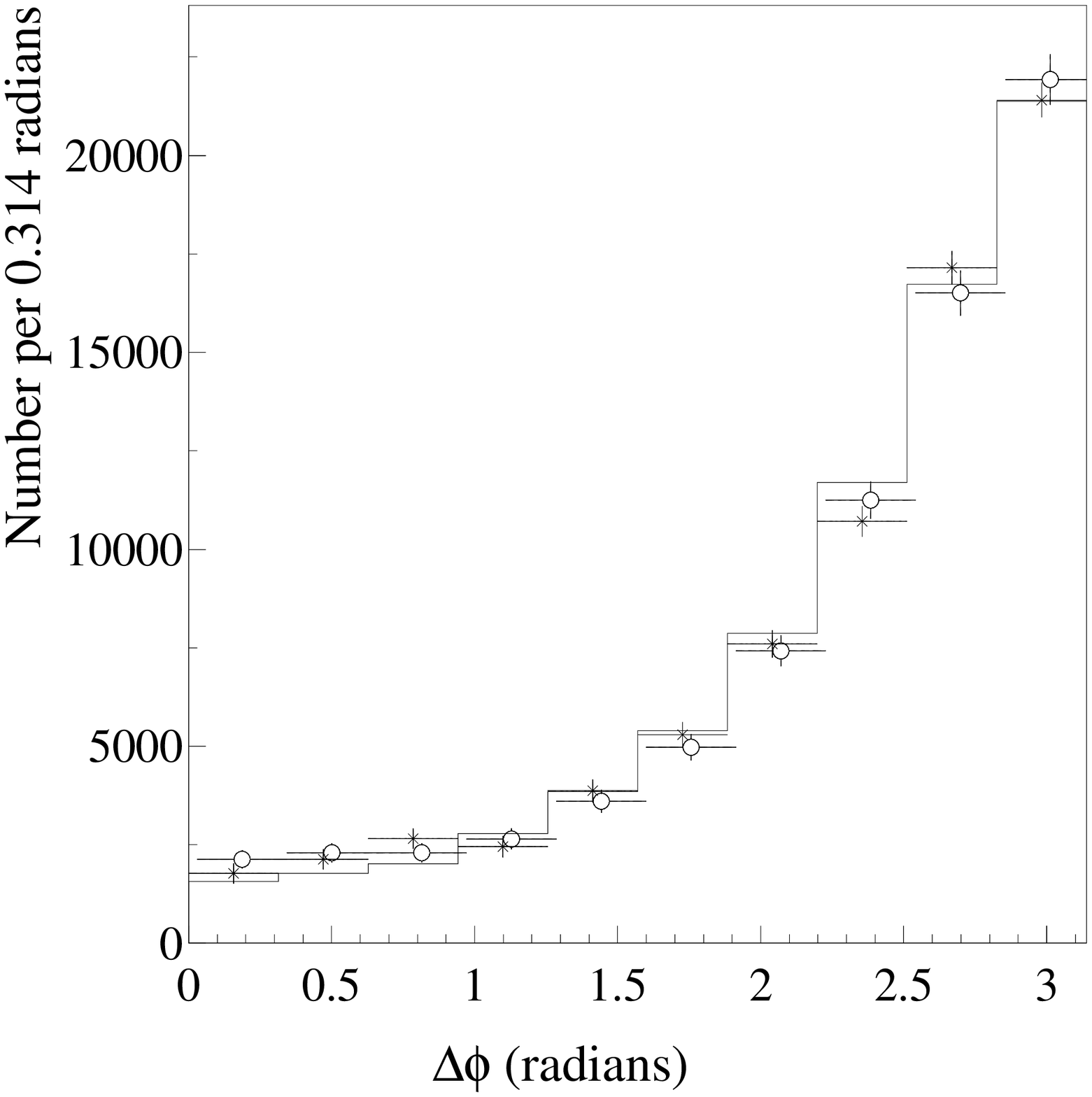}}
\subfigure[]
{\includegraphics[width=0.4\textwidth,clip=true]{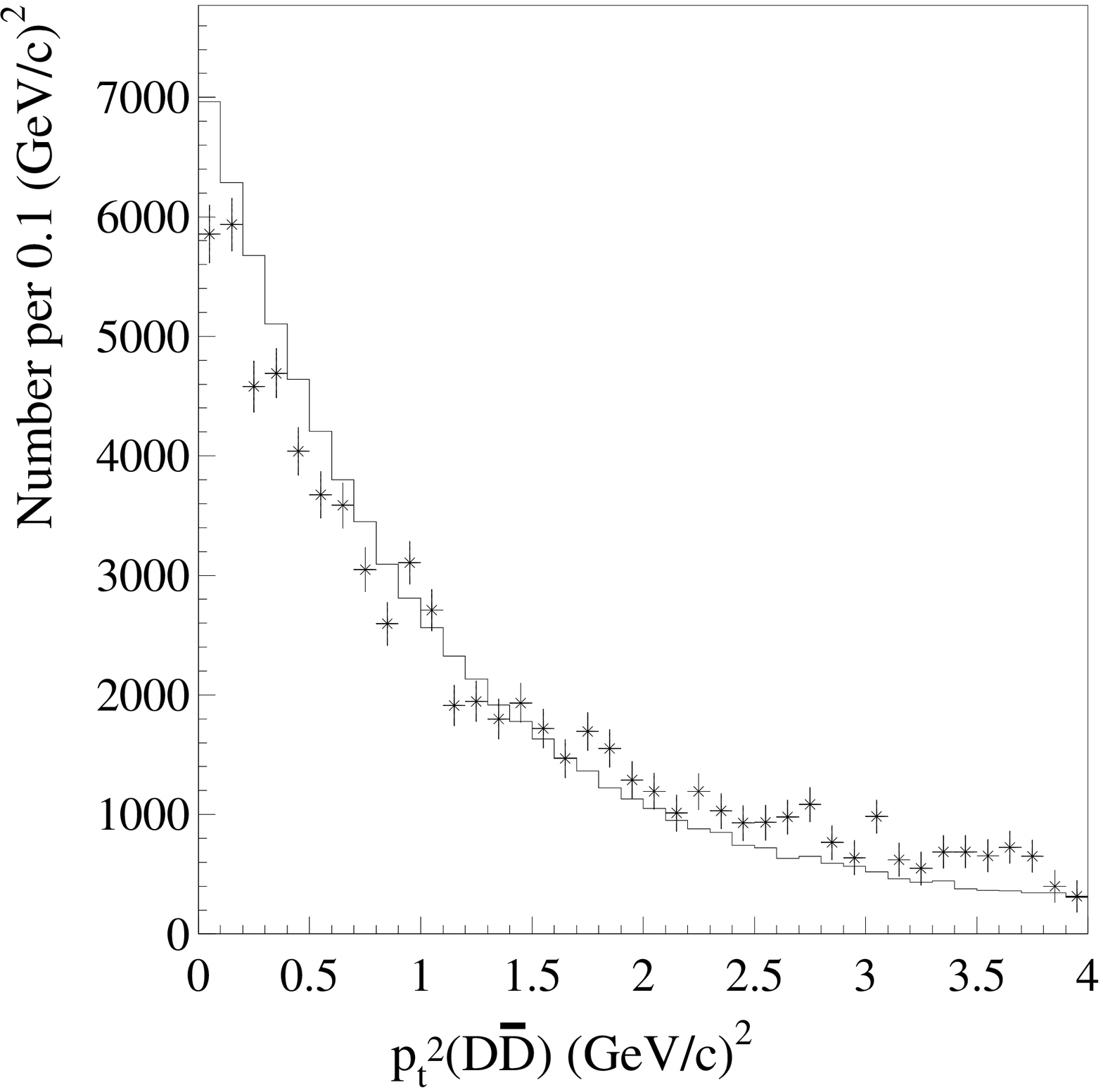}} \\
\subfigure[]
{\includegraphics[width=0.4\textwidth,clip=true]{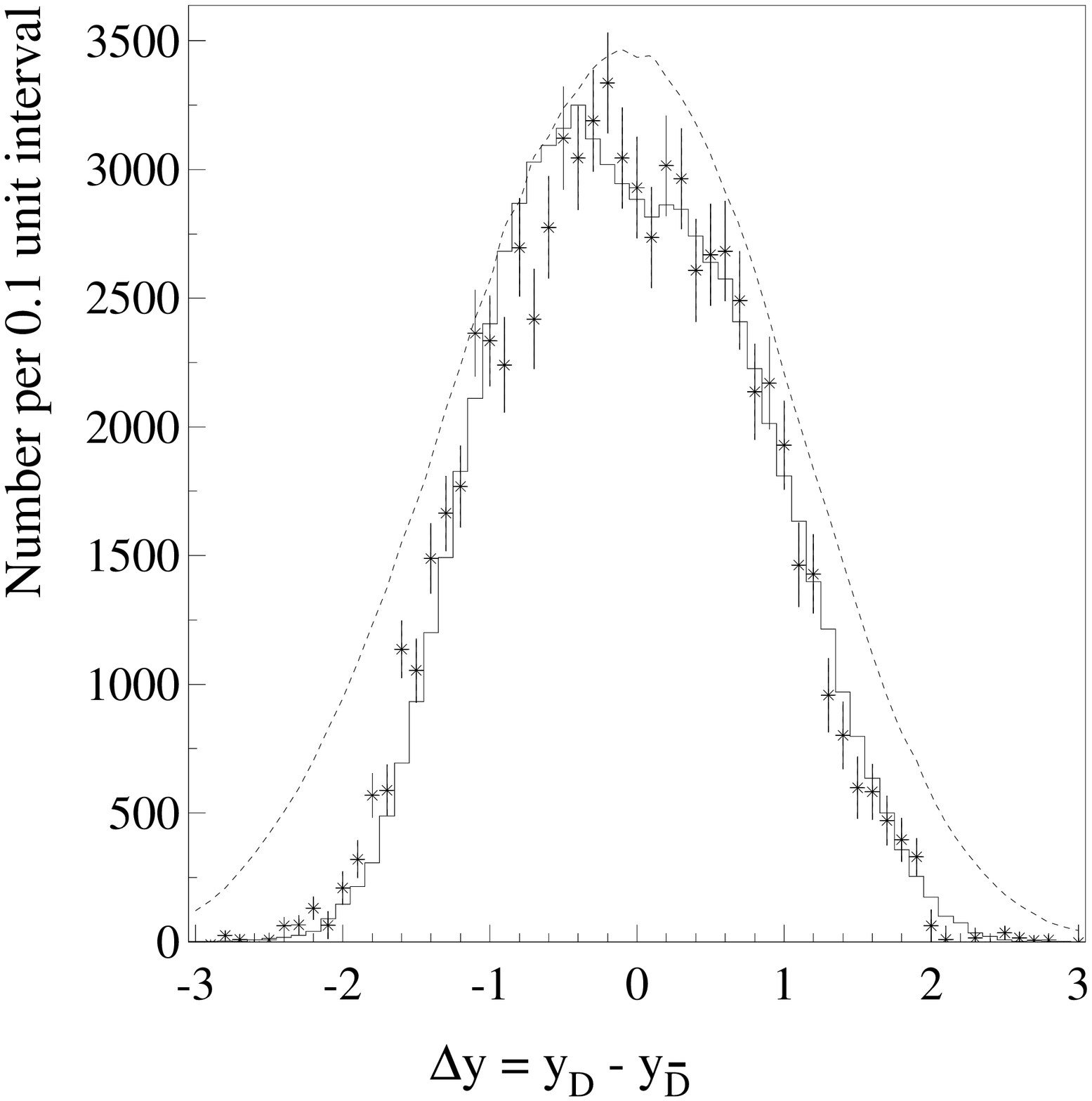}}
\subfigure[]
{\includegraphics[width=0.4\textwidth,clip=true]{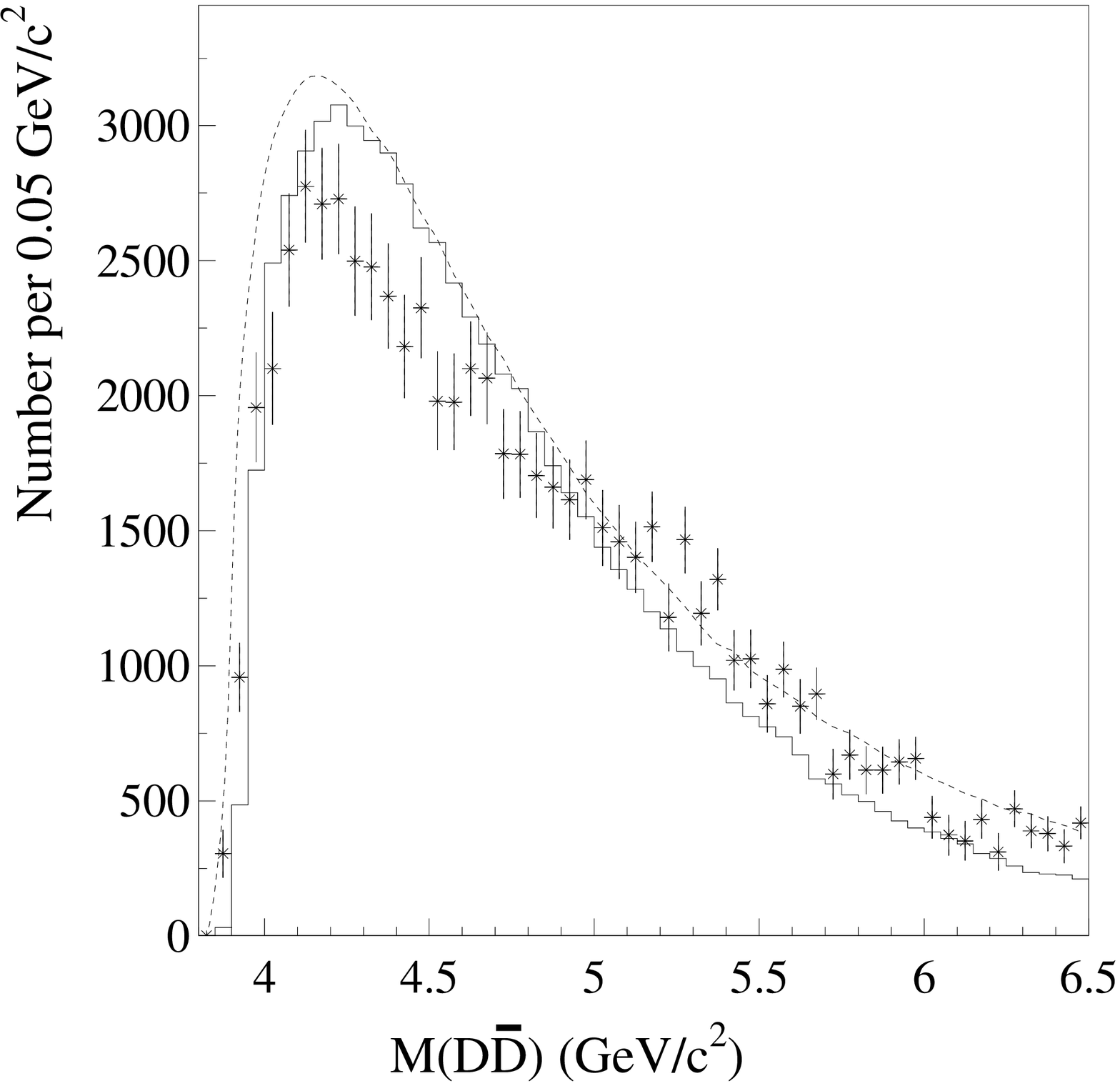}}
\label{figure4}
\caption{
Charm-pair correlations for the partially reconstructed charm-pair sample:
(a) $\Delta\phi$, (b) $p_t^2$ of the $D{\overline D}$ pair, 
(c) rapidity difference ($y_D - y_{\overline D}$), and
(d) invariant $D{\overline D}$ mass for background-subtracted FOCUS data
(data points with error bars) and \textsc{Pythia} 6.203 after detector simulation and
data analysis cuts (solid line). The $\Delta\phi$ distribution for
fully-reconstructed charm pairs (open circles with error bars) is included in
(a) after accounting for resolution broadening (see footnote 10). \textsc{Pythia} parent
distributions (dashed lines with arbitrary normalization) are included in (c)
and (d). The slight distortion (dip) at the peak of the ${\Delta}y$ distribution in (c)
is caused by the \emph{anti-$D^*$} cut described in the text.
}
\end{figure}

To improve comparisons between data and model predictions based on photon-gluon
fusion, we eliminate our lowest multiplicity events by requiring a minimum
number of particles assigned to the primary interaction vertex. We define
$N_{\textrm{primary}}$ as the number of particles assigned to the primary vertex.
With this definition, $N_{\textrm{primary}}$ has a minimum value of two since it
includes the $D$ and ${\overline D}$ mesons (each charm meson counts as a
single particle) in addition to charged tracks assigned to the primary vertex.
To eliminate our lowest multiplicity events we require an $N_{\textrm{primary}}>2$
cut. The cut eliminates features observed in data that are not present in
\textsc{Pythia} 6.203. This is illustrated in Fig.~2a, which shows the
background-subtracted\footnote{The background subtraction procedure assigns
unit weight to $D{\overline D}$ candidates in the signal region in Fig.~1a
($\pm~2\sigma$ about the center of the distribution), a weight of $-^1/_2$ to
candidates in the single $D$ and single ${\overline D}$ sidebands
(four regions defined as $\pm~2\sigma$ about the $D$ axis and $\pm$~4--8$\sigma$
about the ${\overline D}$ axis, and $\pm~2\sigma$ about the ${\overline D}$ axis
and $\pm$~4--8$\sigma$ about the $D$ axis),
and a weight of $+^1/_4$ to
candidates in the four regions where both the $D$ and ${\overline D}$
candidates are 4--8$\sigma$ away from the center of the distribution. The
weight factor of $+^1/_4$ accounts for the over-subtraction of the single-$D$ 
and single-${\overline D}$ backgrounds and the subtraction of random
combinatoric background.}
invariant $D{\overline D}$ mass for mass combinations
with a net charge of zero ($D^+D^-$ and $D^0{\overline D}^0$)
for FOCUS data, and for \textsc{Pythia} events
that have passed through a software simulation of the FOCUS detector and have
survived the event selection procedure described earlier in this paper. The
mass distribution has an enhancement near threshold that is not present in
\textsc{Pythia}. This enhancement is evident for events with $N_{\textrm{primary}}=2$,
especially when we apply additional cuts that remove events with energy
deposited in electromagnetic calorimeters (see inset in Fig.~2a). The enhancement
seems to arise from the diffractive production of $\psi(3770)$ decaying to
$D{\overline D}$, and will be the subject of a future paper (additional
information can be found in conference proceedings~\cite{ichep}). Another
significant difference between data and \textsc{Pythia} is shown in Figs.~2a and 2b. This
is the excess of $N_{\textrm{primary}}=2$ events in data compared to \textsc{Pythia}, some
of which can be attributed to the production of $\psi(3770)$. By eliminating
$N_{\textrm{primary}}=2$ events, we get fairly good agreement for the
$N_{\textrm{primary}}$ distribution in Fig.~2b, which shows the histogram for
\textsc{Pythia} (solid line) normalized to the number of $D{\overline D}$ pairs in the
data (data points with error bars) with $N_{\textrm{primary}}>2$. By eliminating
the $N_{\textrm{primary}}=2$ bin the agreement between data and \textsc{Pythia} is
significantly improved (a slight excess of events with $N_{\textrm{primary}}=3$
persists in the data).

Previous studies~\cite{e791,wa92,e653,na32,wa75,na27,na14,e687}
of charm-pair correlations have presented distributions for
$p_t^2(D{\overline D})$,
the transverse momentum squared of the $D{\overline D}$ pair,
and $\Delta\phi$, the azimuthal angle between the $D$
and ${\overline D}$ momentum vectors in the plane transverse to the beam
direction. These distributions are significant, since $p_t^2(D{\overline
D})~=~0$ and $\Delta\phi~=~\pi$ radians in leading-order QCD, where the
charm-quark pair is produced back-to-back. In QCD these distributions are
broadened by NLO corrections and non-perturbative effects, as illustrated in
references~\cite{frix94x} and \cite{man93}. Photoproduction results from
E687~\cite{e687} have been compared to results from NLO
calculations~\cite{frix98} and \textsc{Pythia} version 5.6~\cite{pyth56}. The E687
comparisons between data and \textsc{Pythia} 5.6 are reproduced in Figs.~2c and 2d, but
with a different normalization to match FOCUS data (shown as open circles with
error bars). The figures show good agreement between FOCUS and E687 data, and a
significant discrepancy between data and \textsc{Pythia}~5.6.

Agreement between FOCUS data and the more recent \textsc{Pythia} 6.203 is significantly
better, but minor discrepancies persist. Figure~3 shows comparisons for
$\Delta\phi$, $p_t^2(D{\overline D})$, rapidity difference defined as
${\Delta}y$~=~$y_D - y_{\overline D}$, and invariant $D{\overline D}$ mass,
$M(D{\overline D})$. FOCUS data are plotted as data points with error bars.
\textsc{Pythia} parent distributions (dashed lines) are shown without acceptance or
resolution effects, so that parent distributions can be compared to the
distributions that are obtained for Monte Carlo events that have survived
detector simulation, event selection and analysis cuts (solid histograms).

Figure~3a shows good agreement for $\Delta\phi$.  There is an enhancement in the
first $\Delta\phi$ bin, which is not present in \textsc{Pythia} and may suggest the
presence of an additional production mechanism.
There is good agreement
for $p_t^2(D{\overline D})$ in Fig.~3b, except that the data tend to have
slightly larger values of $p_t^2(D{\overline D})$. Compared to \textsc{Pythia} 5.6, the
agreement between data and \textsc{Pythia} 6.203 for $\Delta\phi$ and $p_t^2(D{\overline
D})$ is significantly better. Some of the improvement can be attributed to a
larger value for the intrinsic transverse momentum of the incoming partons,
referred to as the $k_T$ kick\footnote{A value of
$<k_T^2>$~=~$(1~\textrm{GeV}/c)^2$ was
introduced with \textsc{Pythia} version 6.135, while previous versions had a value of
$<k_T^2>$~=~$(0.44~\textrm{GeV}/c)^2$},
but a number of other \textsc{Pythia} modifications that
affect these distributions have also occurred over time. Figure~3c shows fairly
good agreement for ${\Delta}y$\footnote{The agreement between data and \textsc{Pythia}
improves slightly for $D$ mesons with larger values of $\ell/\sigma_\ell$,
however a more restrictive $\ell/\sigma_\ell$ cut also reduces the number of
charm-pair events that are available for correlation studies.},
but also shows significant acceptance losses for
$|{\Delta}y| > 1$ (acceptance losses are less severe in the partially
reconstructed charm-pair sample). Acceptance losses are also significant for
large values of $M(D{\overline D})$ in Fig.~3d, but here there is a discrepancy
between data and \textsc{Pythia} for smaller values of $M(D{\overline D})$ where the
acceptance is good.

Figure~4 shows results for the partially reconstructed charm-pair sample,
comparing data (asterisks with error bars) to \textsc{Pythia} 6.203 (solid lines). For
$\Delta\phi$ (see Fig.~4a) we also include a comparison to the distribution that
we obtain for fully reconstructed charm pairs after accounting for resolution
broadening effects\footnote{The $\Delta\phi$ distribution for the
fully-reconstructed sample is obtained by taking the momentum vector of the $D$
or ${\overline D}$ in an event and treating it as the momentum of a $D^*$ that
decays isotropically to a $D^0$ and a pion. The pion momentum vector is then
used to determine $\Delta\phi$ as is done in the analysis of partially
reconstructed charm pair events.}.
This shows that the two samples are in agreement, and that the enhancement that
we observe in the first $\Delta\phi$ bin for fully reconstructed charm pairs
(see Fig.~3a) disappears due to resolution broadening and selection cuts applied
to the partially reconstructed charm-pair sample.
Figures~4a and 4b are both affected by resolution broadening (the
effects are reproduced by our Monte Carlo), and the agreement between data and
\textsc{Pythia} 6.203 is good. As before, the data tend to have slightly larger
values of $p_t^2(D{\overline D})$.

In Figs.~4c and 4d we show results for ${\Delta}y$ and $M(D{\overline D})$, and
include \textsc{Pythia} parent distributions (dashed lines) to show how acceptance
losses in this sample compare to acceptance losses in the fully-reconstructed
sample (see Fig.~3). The partially reconstructed charm pairs are less affected
by acceptance losses, and thus extend the kinematic range of our correlation
studies. The ${\Delta}y$ distributions in Fig.~4c show good agreement,
while the $M(D{\overline D})$ distributions in Fig.~4d exhibit a mismatch
between data and \textsc{Pythia} that is similar to the mismatch that is observed
in Fig.~3d.

\section{Conclusions}

We have extracted two large samples of photoproduced charm-pair events for
studies of correlations between $D$ and ${\overline D}$ mesons. The first
sample consists of more than 7000 fully reconstructed $D{\overline D}$ pairs.
The second sample consists of over $75\,000$ partially reconstructed charm
pairs, where one $D$ meson is fully reconstructed and the other is tagged by a
slow pion coming from a $D^*$ decay. For the fully reconstructed sample we
impose an $N_{\textrm{primary}}>2$ cut to eliminate our lowest multiplicity
events, while the partially reconstructed sample has an implicit cut of
$N_{\textrm{primary}}>2$ due to the presence of the slow pion. The significance
of the $N_{\textrm{primary}}$ cut is that it improves our comparisons to model
predictions based on photon-gluon fusion by eliminating low multiplicity events
in which we observe the production of $\psi(3770)$ decaying to $D{\overline D}$
pairs. The $\psi(3770)$ events, which are not included in \textsc{Pythia}, appear to be
produced diffractively, and will be the subject of a future paper.

The FOCUS results on charm-pair correlations presented in this paper
are in good agreement with previous measurements from experiment E687, which
displayed significant discrepancies compared to an older version of \textsc{Pythia}
(version 5.6). Comparisons of FOCUS data to a more recent version of \textsc{Pythia}
(version 6.203) are significantly better, due to changes in parameters that
affect the modeling of photon-gluon fusion. One notable change that improves
the agreement with data is that the intrinsic transverse momentum ($k_T$) of
incoming partons was
increased
from $<k_T^2>$~=~$(0.44~\textrm{GeV}/c)^2$
to $<k_T^2>$~=~$(1~\textrm{GeV}/c)^2$.
Although minor discrepancies persist when
FOCUS data are compared to \textsc{Pythia}, the modeling of
heavy quark photoproduction is
fairly good for correlations between $D$ and ${\overline D}$ mesons.

\section{Acknowledgments}

We wish to acknowledge the assistance of the staffs of Fermi National
Accelerator Laboratory, the INFN of Italy, and the physics departments of the
collaborating institutions. This research was supported in part by the U.~S.
National Science Foundation, the U.~S. Department of Energy, the Italian
Istituto Nazionale di Fisica Nucleare and Ministero dell'Istruzione 
dell'Universit\`a e della Ricerca, the Brazilian Conselho Nacional de
Desenvolvimento Cient\'{\i}fico e Tecnol\'ogico, CONACyT-M\'exico, and the
Korea Research Foundation of the Korean Ministry of Education.

%
%
%============================  REFERENCES =======================
%
%***** papers and notes

\end{document}